# An AI-enabled Bias-Free Respiratory Disease Diagnosis Model using Cough Audio: A Case Study for COVID-19

Tabish Saeed[*fj], Aneeqa Ijaz[*fj], Ismail Sadiq[‡], Haneya N. Qureshi[*], Ali Rizwan[§], and Ali Imran[*‡]
[*]AI4Networks Research Center, Dept. of Electrical & Computer Engineering, University of Oklahoma, USA
[§] AI4lyf, Pakistan
[‡]James Watt School of Engineering, University of Glasgow, UK
Email: {t.saeed,aneeqa,haneya,ali.imran}@ou.edu

***Abstract*—Cough-based diagnosis for Respiratory Diseases (RDs) using Artificial Intelligence (AI) has attracted considerable attention, yet many existing studies overlook confounding variables in their predictive models. These variables can distort the relationship between cough recordings (input data) and RD status (output variable), leading to biased associations and unrealistic model performance. To address this gap, we propose the Bias-Free Network (RBF-Net), an end-to-end solution that effectively mitigates the impact of confounders in the training data distribution. RBF-Net ensures accurate and unbiased RD diagnosis features, emphasizing its relevance by incorporating a COVID-19 dataset in this study. This approach aims to enhance the reliability of AI-based RD diagnosis models by navigating the challenges posed by confounding variables. A hybrid of a Convolutional Neural Networks (CNN) and Long-Short Term Memory (LSTM) networks is proposed for the feature encoder module of RBF-Net. An additional bias predictor is incorporated in the classification scheme to formulate a conditional Generative Adversarial Network (c-GAN) which helps in decorrelating the impact of confounding variables from RD prediction. The merit of RBF-Net is demonstrated by comparing classification performance with State-of-The-Art (SoTA) Deep Learning (DL) model (CNN-LSTM) after training on different unbalanced COVID-19 data sets, created by using a large-scale proprietary cough data set. RBF-Net proved its robustness against extremely biased training scenarios by achieving test set accuracies of 84.1%, 84.6%, and 80.5% for the following confounding variables - gender, age, and smoking status, respectively. RBF-Net outperforms the CNN-LSTM model test set accuracies by 5.5%, 7.7%, and 8.2%, respectively.**

## I. Introduction

Respiratory diseases (RDs) are among the most common chronic diseases, and a primary cause of morbidity and mortality, imposing an immense burden on health worldwide [1]. COPD, acute lower tract infections, asthma, lung cancer, tuberculosis, and the recent COVID-19 infection are considered the most common acute respiratory diseases [2]. Millions die due to chronic RDs every year because of the lack of timely and accurate diagnosis [3]. The reliable diagnostic tests that are available for the detection of RDs are mostly laboratory-based, expensive, and time-consuming. For instance, a reverse transcriptase polymerase chain reaction (RT–PCR) test specific to SARS-CoV-2 virus is routinely used for reliable detection. This test cannot be used for continuous monitoring, since it can take up to 2 days for determining the diagnosis [4]. Additionally, to rule out the possibility of false negative results, repetitive testing may be required. This underscores the pivotal need for devising alternative methods for rapid, cost-effective, and accurate diagnosis of RDs. Moreover, there exists a dearth of reliable instantaneous screening tools that can detect the RDs at their onset [5]. Such tools are necessary for curbing the spread of the contagious RDs and averting the deterioration of public health.

As a pragmatic solution to this momentous problem, cough-acoustic signals can play a crucial role in monitoring and detecting RD more effectively. Recent surveys [5], [6] indicate that individuals with respiratory illness exhibit distinct features in their acoustic signals, representative of the vocal tracts, which can be extracted. Hence, features extracted from cough can help to continuously monitor and establish the health status of individuals at risk of RDs such as tuberculosis, asthma, COPD, and COVID-19 [2], [5], [7]. Therefore, a smart, accessible, and cost-effective self-monitoring framework for continuous cough monitoring and quick disease diagnosis [7] can be devised by incorporating the recording of cough sounds and implementing AI-based cough processing solutions [8], [9].

Our seminal work on AI-enabled cough-based disease diagnosis has further triggered this interest, where RDs such as COVID-19 can be accurately diagnosed using cough. We demonstrated that it is possible to design advanced Machine Learning (ML) models to: 1) discern cough from non-cough sounds recorded via a smartphone app, and; 2) detect RDs such as COVID-19 infection, bronchitis, bronchiolitis, and pertussis from cough sounds recorded via the same app [10], [11]. In recent years, a large number of independent studies have proposed solutions for digital cough monitoring and collection for timely diagnosis of RDs [12]–[17]. The current literature on the SoTA cough-based RD detection leverages several ML and Deep Learning (DL) algorithms that can classify various temporal, spectral, and statistical cough features, including those that are perceptually indistinguishable to the human ear [5]. The cough-acoustic AI-based RD diagnosis models

[fj] T. S. and A. I. contributed equally to this paper.

include traditional methods like support vector machines [18] for the classification of croup from pneumonia, asthma, bronchiolitis; random forests model used for COVID-19 diagnosis [19]; logistic regression model for the classification of croup and pneumonia [12]; gradient boosting for COVID-19 classification [20], [21]. Several DL models have also been leveraged for cough-based RD diagnosis including CNNs [8], [9] for COVID-19 diagnosis and for croup, pertussis, bronchitis, and asthma classification [22]. Deep neural networks and spiking neural networks have also been used for the classification of COVID-19, asthma, bronchitis, and pertussis with remarkable accuracy [16], [23]. Hybrid, e.g. CNN-LSTM, and ensemble models are also implemented in recent studies for RD diagnosis, achieving high accuracy [17], [24]–[31].

In the rapidly advancing field of AI-enabled automated cough sound monitoring and digital disease diagnosis, remarkable performance metrics have been achieved as evidenced by several notable studies [8], [10], [14], [32]–[36]. However, a critical concern remains unaddressed in the majority of these studies: the potential impact of confounding variables and data biases on the performance of the AI models they employ. These models tend to overestimate their classification performance and overfit to the training data biases, while falling short in terms of proper validation and generalization to unseen data. A recent investigation by the University of Cambridge in the UK underscored this glaring deficiency within a substantial body of research dedicated to accurately detecting and diagnosing COVID-19, highlighting the oversight of confounders in the evaluation of AI frameworks [35]. The root issue is that confounding variables can distort the apparent relationship between input features and diagnostic outcomes, leading to erroneous predictions [37], [38]. For instance, studies aiming to distinguish individuals with a disease from healthy controls often face the challenge of dealing with a substantial age difference between the two groups. In such cases, the AI model may inadvertently learn associations primarily influenced by age disparities rather than the genuine disease-related biomarkers and features, thus severely hindering its ability to generalize its findings. Importantly, these confounding factors can include a range of biases, such as those related to age, gender, race, and medical history, all of which can introduce systemic inaccuracies and pose potential threats to equitable healthcare assessments outcomes. Addressing these challenges is crucial for improving the reliability and fairness of AI models in the context of disease diagnosis and monitoring. The findings of our study highlight the importance for future studies to consider accounting for the effects of the confounding variables, similar to RBF-Net, so the reported results are a realistic representation of classification expected in a real-world scenario.

To address this limitation in the existing studies on cough-based diagnosis, we propose an end-to-end generalized RD Bias-Free Network (RBF-Net) and evaluate its efficacy on COVID-19 dataset. To the best of authors' knowledge, this is the first study that proposes a framework that is robust to the confounding variables for COVID-19 diagnosis; thus, providing realistic and generalized performance. The proposed RBF-Net framework contains a bias predictor module which helps in identifying features from the cough recordings that are statistically invariant to confounding effects and mainly characterized by the effects of COVID-19, using an adversarial learning technique [39].

The contributions of this work are summarised as follows:

- In contrast to the majority of the previous studies that rely on the crowd-sourced cough audio databases for training AI models, this study curated a cough data set, containing COVID-19 infection status. For each participant, the curated data set includes cough recordings, tagged with reliable RT-PCR information, collected in a clinical setting. Hence, the data set used has extremely reliable ground truth labels, resulting in the accurate training of RBF-Net.
- To demonstrate the impact of confounding variables, we train a SoTA DL model on different splits of biased training scenarios from the cough data set based on gender, age, and smoking status. Moreover, we present an insightful analysis on how model performances are often overestimated due to the underlying biased distribution of the training data and the use of cross-validation technique.
- To overcome the impact of biases, we present RBF-Net that learns features, from the cough recordings, that are impacted by the COVID-19. We perform a comparative analysis of the existing SoTA CNN-LSTM model with RBF-Net and demonstrate the improvement achieved by the proposed RBF-Net in terms of different performance metrics.

The remaining contents of this paper are organized as follows: Section II discusses the details of the cough data acquisition and its pre-processing. Section III presents the proposed RBF-Net architecture, Section IV describes the methodology adopted for the study. The results for classification on data with various biases using the proposed RBF-Net and the existing SoTA CNN-LSTM are explained in Section V. Section VI discusses the impact of our work, future clinical deployment, and acknowledges some limitations of this work. Finally, the conclusion for the study is given in Section VII.

## II. Cough Data Acquisition and Pre-processing

We have collected a corpus of high-fidelity audio data containing cough acoustics of normal and COVID-19 diagnosed patients. The notable feature about the data set is that rather than being collected through crowd-sourcing, it was curated to have a valid tagged RT-PCR test result for each sample. The audio data sample acquisition was performed during the time period of Apr - Oct 2020, in collaboration with Dow Medical College, Pakistan. For this research, the cough data samples were recorded from the subjects through our in-house developed AI4COVID app [40], under the supervision of trained nurses, using one smartphone model to avoid the impact of device variability. Each participant recorded multiple coughs in a recording sample, with each recording duration varying from 3 to 12 seconds. An informed consent was obtained from each participant prior to acquiring the cough data. The guidelines to interact with the potential COVID-19 patients recommended by the WHO were strictly followed at all stages of the cough data collection. For instance, the healthcare professionals wore personal protective equipment (PPE), and followed a protocol for the smartphone disinfection, before and after the sample was recorded. In total, the data is collected from 1094 participants with positive RT-PCR test results, labelled as COVID-19 positive, and 3761 participants with negative RT-PCR test results, labelled as normal. In addition, the anonymity of the users was preserved at all stages during the data collection.

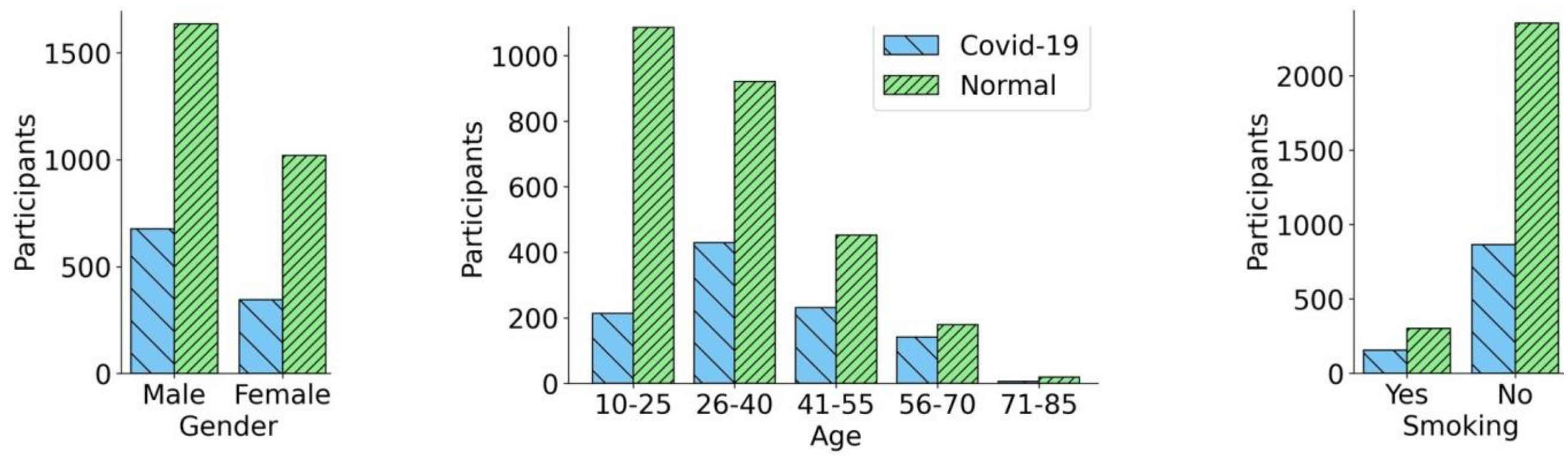

Fig. 1: Bar plots representing the detailed demographic and smoking status statistics for the selected participants after the process of data cleaning. A total of 1022 COVID-19 positive participants and 2656 normal participants are finalized.

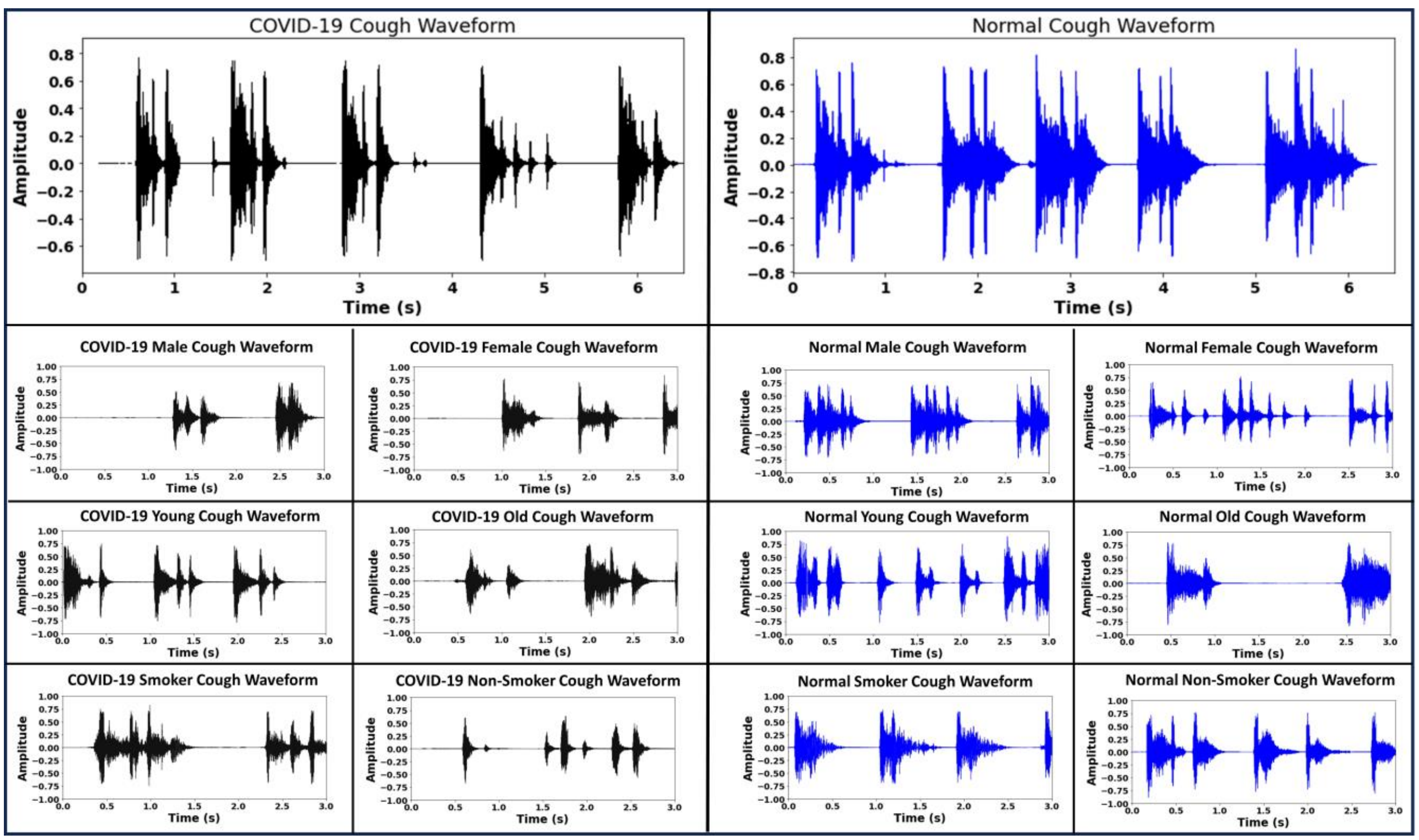

Fig. 2: Sample waveforms for each target class, i.e. COVID-19 and normal. The waveforms contain information regarding the gender, age and smoking status, that can cause a classifier to learn inaccurate representation for a cough recording with respect to a target disease label.

Once the data acquisition process was complete, we performed the cough sound pre-processing steps, including noise removal, using the Audacity software [41]. The mono-channel cough data is sampled at 44.10 kHz, before being stored as a pulse-code modulation (PCM) WAV file. Silent periods at the beginning and at the end of each cough recording were clipped out. After the initial pre-processing, only the cough recordings that were longer than 2 seconds of duration were considered for further analysis. Thus, at the completion of the pre-processing step, we had a total of 1022 COVID-19 positive samples and 2656 normal samples. Fig. 1 provides a summary of the demographics for the individuals constituting the cough audio data set. It illustrates the number of COVID-19 and normal participants with respect to their demographics, including gender, age, and smoking status. It can be observed from the Fig. 1 that both normal samples and COVID-19 positive samples are skewed towards the male gender, i.e., the male cough samples are almost twice as numerous compared to the female samples. Similarly the age of the subjects in the data set fall in a broad age range from 10 to 85 years, with a significant number of the cough samples belonging to the younger and middle aged population, 18-50 years old. A small fraction of the participants are smokers, 865 subjects who tested COVID-19 positive and 2355 normal participants were non-smokers. On the other hand, there were a total of 156 COVID-19 positive smoking participants and 301 normal smoking participants. Fig. 2, demonstrates the sample waveforms from the COVID-19 and normal classes, respiratory disease affects the lungs and results in changes in the acoustic signature of the cough sound. These changes are not always clearly identifiable on inspection. Deep neural networks trained on audio data comprising of coughs from respiratory disease can learn to identify cough acoustic features characterizing the respiratory disease [5]. In addition to disease status the cough waveforms have information like gender, age and smoking habits encoded within them that can lead the classifier to learn inaccurate representations for the disease labels. By incorporating the bias-free mechanism, the RBF-Net learns to disassociate the effects of the confounding factors like gender, age and smoking status from the disease status when classifying the cough recording, further details are given in section IV.

## III. RBF-Net Architecture

In this section, we present the detailed architecture of the proposed RBF-Net with focus on COVID-19 classification. CNN-LSTM model forms the main skeleton of the proposed framework architecture. The CNN-LSTM has proven to be the SoTA spectrogram-based COVID-19 classification DL model and has been widely used for the classification tasks by the research community [17], [24]–[31]. We use this SoTA DL technique as both the benchmark for our framework and also the building block for the feature encoder module of the proposed RBF-Net. The architectural details of the implemented CNN-LSTM model are shown in Fig. 3.a).

The CNN-LSTM architecture can primarily be divided into two blocks. The first block (feature encoder) uses a CNN architecture which receives grayscale spectrograms, constructed from cough recordings, as an input of shape 224x224. Then, the most relevant and informative features are extracted by the convolutional layers. These features are converted to the feature maps, which are passed on to the LSTM block, where the deep features that have the high temporal correlation are selected to capture the more useful patterns. In the second block (COVID-19 classifier), a simple fully connected layer is used for the feature learning and COVID-19 classification. Both of these blocks are trained through the COVID-19 classification loss ($L_c$) which is chosen to be the binary cross-entropy loss. The $L_c$ is backpropagated in a manner that the model parameters of the feature encoder ($\partial_E$) and COVID-19 classifier ($\partial_C$) are tuned to minimize it.

Although this model works extremely well in learning the differences between the two classes in the target variable (COVID-19 or normal) as shown in the Results section, it can not nullify the impact of the confounding variables. This model is prone to be affected by the impact of biases in its learned features; thus, the model cannot be truly representative of COVID-19's impact on the cough sounds. To address the underlying biases and confounding variables in the data distribution, we make modifications to the CNN-LSTM inspired by the recent work in the machine learning fairness schemes [39]. An additional bias predictor component is attached to the network architecture that helps the encoder module to decorrelate the extracted feature vector from the effects of confounding variables, as evidenced by the classification results for the RBF-Net framework in the Results section. This decorrelation process is based on training the feature encoder through an adversarial learning technique similar to the conditional GAN (c-GAN). In this iterative training process, bias predictor aims to predict the bias value from the feature vector created from conditioned subset of normal samples from the input and then have an adversarial impact on the encoder to learn the features that are bias-free. In this manner, the RBF-Net framework learned features conditionally independent of the biases and carrying useful information for COVID-19 classification. Thus, the overall goal to create a bias-free and generalizable RD classifier for large-scale clinical deployment can be realized.

The RBF-Net architecture is composed of three key blocks, depicted in Fig. 3.b), each undergoing distinct training phases. Initially, the COVID-19 classifier block focuses on precise disease classification, dynamically updating its model parameters $\partial_C$. The gradient, propagated with respect to the COVID-19 classification loss $L_C$, refines the encoder parameters $\partial_E$, ensuring feature extraction that minimizes $L_C$. In the subsequent phase, with frozen encoder parameters $\partial_E$, the bias predictor module is trained. This involves updating parameters $\partial_B$ to minimize bias prediction loss $L_B$, determined by either inverse mean squared error or inverse binary cross-entropy based on the bias variable. In the final training step, with $\partial_B$ frozen," the feature encoder undergoes training. Adversarial back-propagation of gradients from $L_B$ fortifies $\partial_E$ to extract features maximizing $L_B$, establishing a min-max game that cultivates bias-invariant features crucial for COVID-19 classification. This strategic training paradigm equips RBF-Net to accurately classify COVID-19 without succumbing to the influence of underlying biased data distributions.

## IV. Methodology

### A. Creation of Cough Spectrograms

In this first stage of the RBF-Net framework, the cough recordings are mapped onto the spectrograms, that depict the spectro-temporal correlations in the audio signal as image. The motivation to use the spectrograms stems from the fact that the spatio-temporal features have the ability to provide comprehensive description of the respiratory sounds (lungs, wheeze, crackles, cough etc.) [42], [43] and to train the deep neural networks [44], [45]. The spectrogram is a two dimensional time-frequency representation of the one dimensional cough acoustic samples. The COVID-19 coughs and normal coughs are transformed into the respective time-frequency representations. A spectrogram divides the time domain into several regions. The x-axis characterizes the time domain while the y-axis represents the frequency domain of the signal. The spectrogram is contructed by a short-time fourier transform (STFT) with short frame size of 25 ms and stride of 10 ms [46]. Then a Hamming window of length 2048 is employed that divides the acoustic signal into the frames. To obtain the spectrogram, a 128-point fast fourier transform (FFT) is implemented on each respective frame. In order to derive the spectrogram final representation, we used the magnitude square of the STFT coefficients. Finally, for achieving the computational efficiency, the spectrogram coefficients are log transformed to introduce the compressive nonlinearity and linearly downsampled onto a 224x224 matrix for cough audio signal.

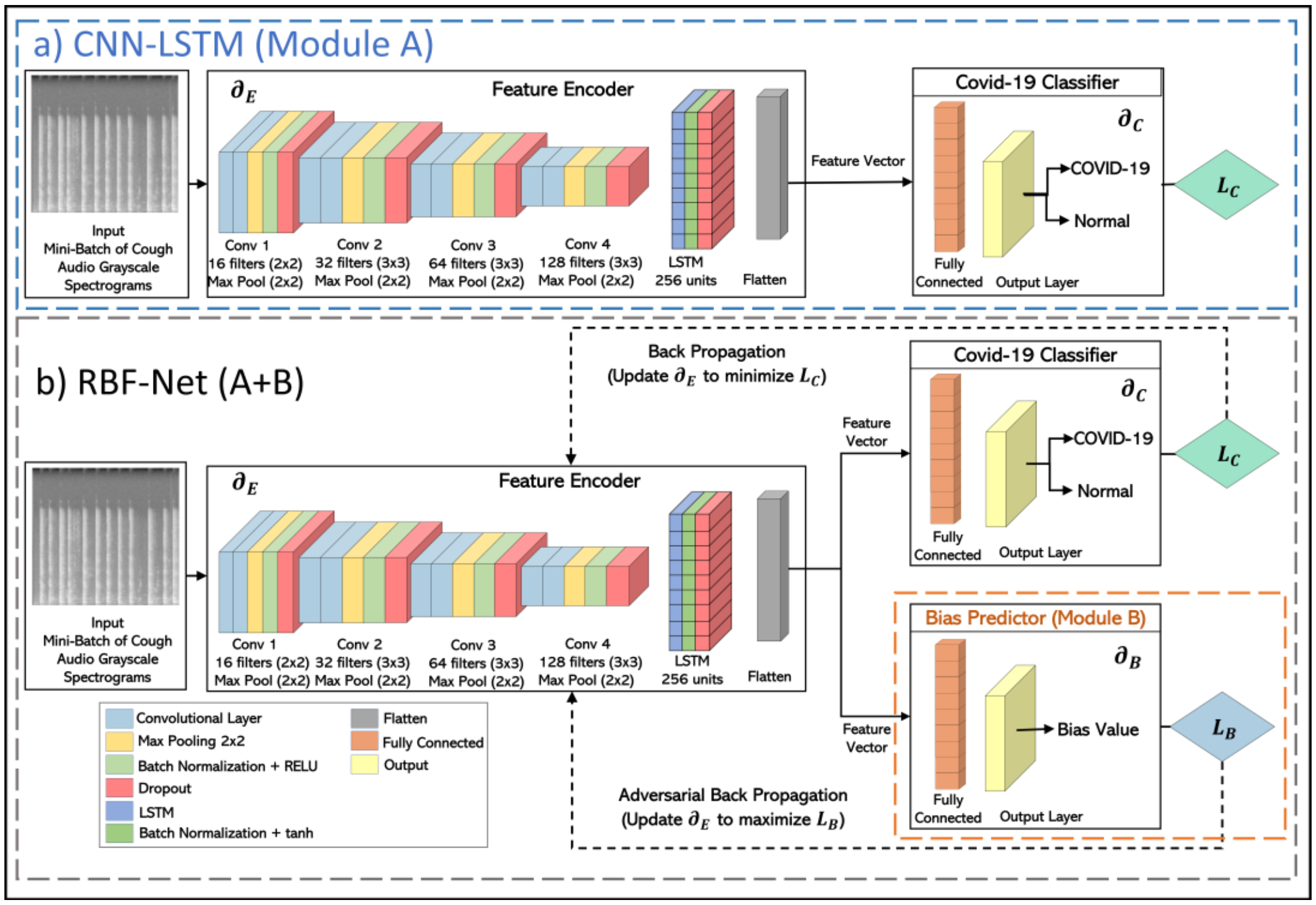

Fig. 3: **Overview of the architecture of the RBF-Net framework**. a) CNN-LSTM model architecture is shown which consists of feature encoder (convolutional and LSTM blocks) and a COVID-19 classifier (fully connected layers with ReLU activation). b) In the RBF-Net, we have an additional bias predictor module (fully connected layers with ReLU activation) that predicts bias from the feature vector. The losses, $L_C$ and $L_B$ are back-propagated to train the encoder through a min-max game similar to c-GAN, so that the extracted features are invariant to the confounding variables.

### *B. Biased Training Data Generation*

To demonstrate how the biases impact the performance of SoTA DL models and to best evaluate the robustness of RBF-Net framework, we have created multiple training data sets with various biases. In these training data sets, different bias factors were induced pertaining to the following confounding variables: gender, age, and smoking status. In each of these sets of training data, the number of COVID-19 positive and normal class participants were kept equal, to properly address the class imbalance problem. An eminent aspect of this work is that after training models using these sets of biased training data, we also create sets of unseen testing data for their evaluation. The unseen testing data sets were created such that they have a balanced number of participants with respect to each of the confounding variables.

*1) Gender Bias:* We create a gender biased training scenario where the COVID-19 positive rate greatly differs in both male and female gender groups. In this training data set, we have 925 participants in each of the classes, i.e., COVID-19 positive and normal. Although, the total number of participants is balanced in both of the classes, the difference lies in the number of male and female participants in each of the class (gender bias). As shown in Fig. 4, COVID-19 training class comprises of a greater number of male participants compared to the female participants. On the other hand, normal training

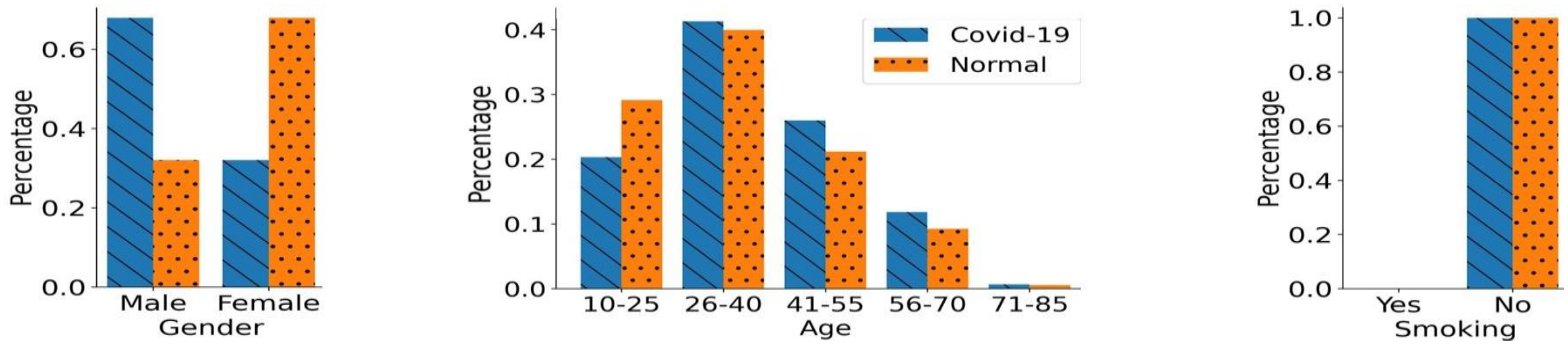

Fig. 4: **Distribution of the Gender Biased Training Data**. In this data set, there is an over-representation of male participants in the COVID-19 class, and there is over-representation of females in the normal class. The age distribution for both the classes is identical. Only non-smoking participants are chosen for the creation of this training data set.

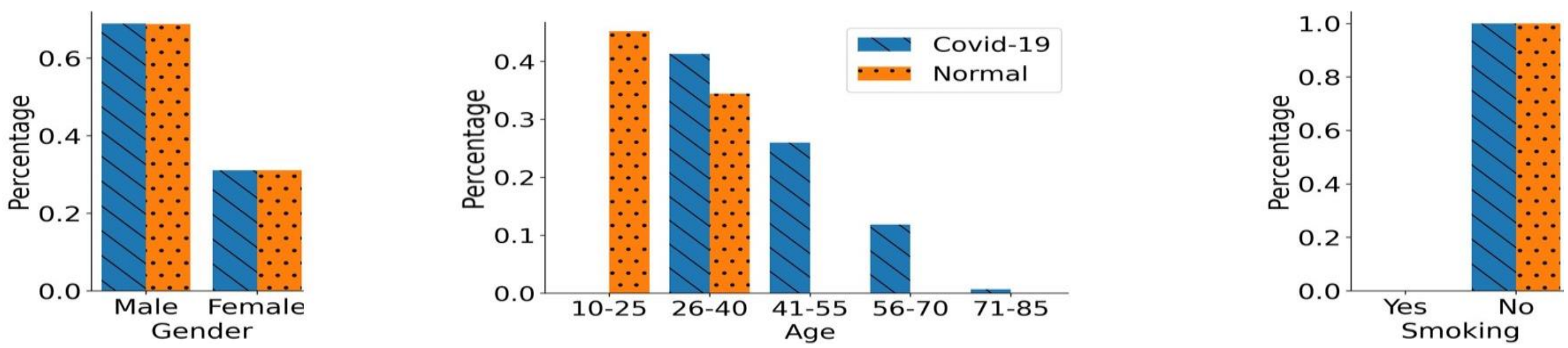

(a) Age Biased (Group 1) Training Data

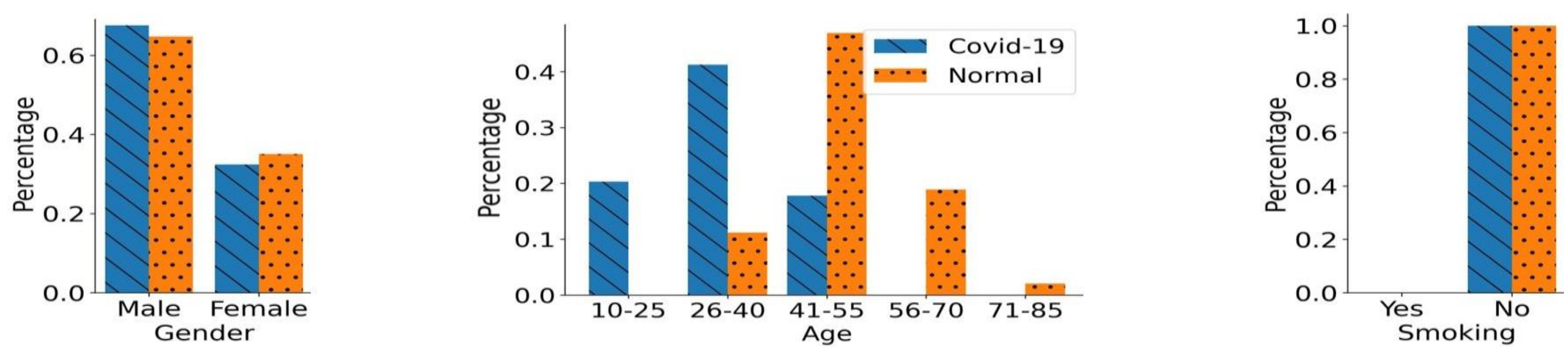

(b) Age Biased (Group 2) Training Data

Fig. 5: **Distribution of the Age biased Training Data Groups. In age biased (group 1), there is an over-representation of relatively younger population (aged under 40) in the normal class compared to the COVID-19 class. In age biased (group 2), there is over-representation of relatively older population in the normal class (aged above 40) compared to the COVID-19 class. The gender distribution for both the classes is identical in these two age biased groups. Only non-smoking participants are chosen for the creation of both of the training data sets.**

class has a greater percentage of female participants. Thus, there exists a bias pertaining to the gender variable in the training data set. Simultaneously, we keep the impact of age and smoking status confounding variables balanced in both of the classes as shown in Fig 4. We have roughly the same number of COVID-19 and normal participants in each of the age bracket. Moreover, we only keep the participants with non-smoking status in both of the classes. Therefore, the bias in the training data set is only induced through gender and not through age and smoking status.

We also created a corresponding gender-based unseen balanced testing data set using the total collected data. In this testing data set, we have 100 participants in each of the classes. We have a balanced number of males and females in both of the classes, i.e., 50 males and 50 females in both COVID-19 and normal classes. Additionally, we keep a similar distribution in this testing data set as the gender biased training data set in terms of age and smoking status variables.

*2) Age Bias:* To further study the impact of another unique confounding variable on the performance of DL models, we created the age biased training data sets. For a more generalised evaluation, we created two age biased training data sets, i.e., age biased group 1 and age biased group 2. In both of the groups, we have 765 participants in each COVID-19 and normal class. The demographics of both of the age bias training data sets are shown in Fig. 5. In group 1, the normal participants were chosen to be from a relatively younger population i.e., aged under 40 years. At the same time, COVID-19 class in group 1 includes participants from a relatively older population. On the contrary, group 2 has different age distributions in both of the classes compared to group 1. In group 2, normal class participants were chosen from an older population i.e., aged above 40 years. Simultaneously, COVID-19 class participants were chosen from a relatively younger population as shown in Fig. 5. In both of these groups, the entire bias is induced only through the age groups. There is no additional bias induced through gender as the number of male and female participants is approximately equal in each of the classes across both groups. Moreover, only non-smoking participants were chosen in these data sets to nullify the impact of smoking status. Apart from these two training data sets, we also created a corresponding age-based unseen balanced testing data set which includes 100 participants in each of the classes. Both of these classes in the testing data set have identical age distribution; thus, having no underlying age bias. Moreover, the testing data set has the same gender and smoking status distribution as the training data sets. Therefore, overall there is no underlying bias in the unseen balanced testing data set.

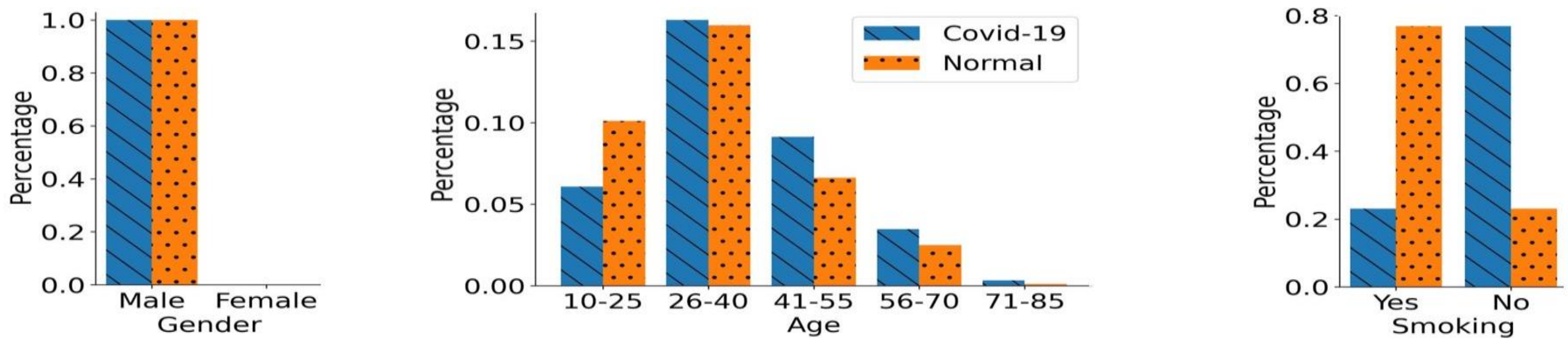

Fig. 6: **Distribution of Smoking Status Biased Training Data**. In this data set, there is an over-representation of non-smoking participants in the COVID-19 class, and there is over-representation of smoking participants in the normal class. The age distribution for both the classes is identical. Only male participants are chosen for the creation of this training data set.

*3) Smoking Status Bias:* For further evaluation of the impact of another distinctive bias on the DL models, smoking status biased training data sets are also created. Smoking status biased training data contains 350 participants in both COVID-19 and normal class. Though, the total number of participants is balanced in both of the classes, the difference lies in the number of smoking and non-smoking participants in each of the classes. As shown in Fig. 6, COVID-19 training class comprises of greater number of non-smoking participants compared to the smoking participants. On the other hand, normal training class has a greater percentage of smoking participants. Thus, there exists a bias pertaining to the smoking status variable in the training data set. Simultaneously, we keep the impact of other two confounding variables balanced in both of the classes as shown in Fig. 6. We have approximately the same age distribution in both classes (COVID-19 and normal). Another important aspect of this smoking biased data set is that as the majority of the original cough database comprises of more male participants compared to the female participants, we only kept the male participants in both of the classes of the smoking biased training data. Therefore, the bias in the training data set is only induced through smoking status and not through gender and age. Similar to the other gender and age confounding variables, we created an unseen balanced testing data set for the smoking status as well. In this data set, the number of non-smoking and smoking participants are balanced across both of the classes. At the same time, this testing data set had the same demographic distribution as the smoking status biased training data, which means that there is no innate bias in the smoking status-based unseen balanced testing data.

TABLE I: Optimal Training Parameters for RBF-Net

| **Hyperparameters** | **Best Value** |
|---|---|
| Number of Convolutional Blocks | **4** |
| LSTM Units | **256** |
| LSTM Activation | **tanh** |
| CNN Activation | **ReLU** |
| Epochs | **1000** |
| Optimizer | **ADAM** |
| Batch Size | **256** |
| Learning Rate | **0.0001** |

## *C. Model Training*

We trained the CNN-LSTM and the RBF-Net models on the data distributions described in Section IV-B, to analyze the impact of bias in the data distribution on model performance. Both models are implemented in Python 3.6 and Tensorflow 1.21. The training process is accelerated through a high performance computing cluster integrated with Nvidia V100 Tensor Core GPUs.

To achieve maximum possible accuracy and generalization, we performed extensive hyper-parameter tuning on both models. The main purpose is to optimize the values of model parameters that are not learned during training, but rather set before the training process. In case of CNN-LSTM and the RBF-Net, the hyperparameters that are searched include the learning rate, epochs, batch size, optimizer, and the number of layers/nodes. Hyperparameter optimization was performed by creating small validation data sets having the same data distribution as the biased training data sets. We used these validation sets and the AutoML optimization tool [47] for tuning each model individually. The sensitivity and specificity metrics of the validation sets after the $100^{th}$ epoch of training are used to identify the optimal hyperparameter configuration. Once these hyperparamters are set, we initiated the training process for both of the models. The best combination of parameters have been listed in Table. I. These CNN-LSTM and RBF-Net models are individually trained using the optimal hyperparameter configuration on each of the biased training set followed by their performance being evaluated on the respective unseen testing sets for a fair comparison.

TABLE II: **Performance comparison on the gender biased training data:** Performance evaluation for CNN-LSTM and RBF-Net model in terms of accuracy, specificity, sensitivity, F-1 score, and ROC-AUC.

| | | Model | Accuracy | Specificity | Sensitivity | F-1 Score | ROC-AUC |
|---|---|---|---|---|---|---|---|
| Gender Bias | Cross-Validation | CNN-LSTM | 0.890 | 0.913 | 0.866 | 0.892 | 0.893 |
| | Unseen Testing Data | CNN-LSTM | 0.787 | 0.860 | 0.715 | 0.785 | 0.787 |
| | | RBF-Net | 0.841 | 0.887 | 0.796 | 0.845 | 0.846 |

TABLE III: **Performance comparison on both of the age biased training data sets (Group 1 and Group 2):** Performance evaluation for CNN-LSTM and RBF-Net model in terms of accuracy, specificity, sensitivity, F-1 score, and ROC-AUC.

| | | Model | Accuracy | Specificity | Sensitivity | F-1 Score | ROC-AUC |
|---|---|---|---|---|---|---|---|
| Age Bias Group 1 | Cross-Validation | CNN-LSTM | 0.887 | 0.908 | 0.867 | 0.894 | 0.892 |
| | Unseen Testing Data | CNN-LSTM | 0.774 | 0.843 | 0.706 | 0.775 | 0.776 |
| | | RBF-Net | 0.845 | 0.888 | 0.801 | 0.845 | 0.846 |
| Age Bias Group 2 | Cross-Validation | CNN-LSTM | 0.884 | 0.901 | 0.867 | 0.884 | 0.881 |
| | Unseen Testing Data | CNN-LSTM | 0.756 | 0.825 | 0.687 | 0.757 | 0.756 |
| | | RBF-Net | 0.818 | 0.863 | 0.774 | 0.819 | 0.821 |

Another crucial aspect of training RBF-Net lies in its convergence and stability. Since RBF-Net fundamentally mimics the c-GAN training scheme for classification of COVID-19 and normal cough sounds, the convergence and stability of the model had to be ensured. The Feature encoder of RBF-Net is trained through a min-max game where it aims to accurately predict COVID-19 while being invariant to the underlying biases. For this reason, the optimal learning rate is very small to avoid fluctuations in the objective loss functions or accuracy metric while training. Moreover, RBF-Net is trained for 1000 epochs during which the performance metrics during the training and validation process start to stabilize with minimal fluctuations. Once these metrics reach a consistent steady state after approximately 650 epochs, the RBF-Net effectively converges. While training on all training scenarios, the RBF-Net training reaches stability and converges for effective COVID-19 prediction without being impacted by confounding variables.

## V. RESULTS

We trained the CNN-LSTM model on all of the biased training data sets to analyze the impact of confounding effects exist in the data distribution on this particular SoTA DL model. We utilized 10-fold cross-validation (CV) on the CNN-LSTM model; the data sets are divided into 10 equal folds. In each iteration of CV, one of these folds is used as a validation set, while the other nine folds are used for the training. This process is repeated 10 times, with each fold being used as the validation set once. During each iteration of the CV, we measured the accuracy, specificity, sensitivity, F1-score, and ROC-AUC for the model performance on the validation set. After all iterations are completed, the average of each performance metric across all iterations is calculated and reported. After this CV technique, the CNN-LSTM model architecture is trained on the biased data sets and its performance is evaluated on the respective unseen balanced testing data sets to gauge if the unrealistic experiment design leads to an inaccurate inflation of the model performance. Furthermore, in order to evaluate the efficacy of the RBF-Net framework in learning the features that are purely extracted through difference in cough spectrograms caused by the impact of COVID-19 respiratory disease, we trained it on the biased training data sets as well. We evaluated the performance of RBF-Net on the same unseen balanced testing data sets and performed a comparative analysis with the CNN-LSTM model. In the subsequent subsections, we evaluate the performance of both CNN-LSTM and RBF-Net on each of the biased training data sets individually.

### A. Performance on Gender Biased Training Scenario

The performance of CNN-LSTM model in COVID-19 detection is inaccurately inflated under the influence of innate gender bias in the training scenario. From the results obtained through CNN-LSTM model using CV technique to the results obtained through CNN-LSTM on the unseen testing data set, a major decline in performance metrics is observed as reported in the Table II. The obtained accuracy drops from 0.890 to 0.787. This clearly indicates that the CV technique leads the CNN-LSTM model to learn the features impacted by the gender bias. Thus, the performance of the CNN-LSTM model in accurately diagnosing COVID-19 is highly overestimated. Another vital observation lies is the difference in the obtained specificity (0.860) and sensitivity (0.715) of the CNN-LSTM model on the testing data set. Since the samples from the normal class participants are over-represented by the female participants in the training data, the model tends to treat female participants as normal participant. On the other hand, the results of the RBF-Net model, when evaluated on the unseen testing data set, demonstrated a significant improvement over the CNN-LSTM, as measured by the five different performance metrics. We have obtained an overall accuracy improvement of more than 5% through the RBF-Net model. The difference between the obtained specificity (0.887) and sensitivity (0.796) is also diminished in the RBF-Net model; thus, it is better suited to mitigate the effect of gender bias in the training scenario.

TABLE IV: **Performance comparison on the smoking status biased training data:** Performance evaluation for CNN-LSTM and RBF-Net in terms of accuracy, specificity, sensitivity, F-1 score, and ROC-AUC.

| | | Model | Accuracy | Specificity | Sensitivity | F-1 Score | ROC-AUC |
|---|---|---|---|---|---|---|---|
| Smoking Bias | Cross-Validation | CNN-LSTM | 0.862 | 0.881 | 0.844 | 0.867 | 0.862 |
| | Unseen Testing Data | CNN-LSTM | 0.723 | 0.750 | 0.694 | 0.727 | 0.723 |
| | | RBF-Net | 0.805 | 0.836 | 0.774 | 0.811 | 0.805 |

### B. Performance on the Age Biased Training Scenarios

Similar to the gender biased data, the performance of CNN-LSTM model in COVID-19 detection is overestimated under the presence of underlying age distribution in the training scenario. Using 10-fold CV, CNN-LSTM model obtains an average accuracy of 0.887 and 0.884 in both of the age bias groups, respectively as shown in Table III. On the other hand, when the CNN-LSTM model is trained on the same training data sets and evaluated on the unseen balanced testing data set, a substantial decline in the performance metrics such as accuracy, specificity, sensitivity, F1-score, and ROC-AUC is observed. It achieves an unseen testing accuracy of 0.774 and 0.756 when trained on both groups, respectively. As shown in the previous section, age bias group 1 has an over-representation of a relatively younger population (aged under 40 years) in the normal participants. This leads the model in treating younger participants as normal samples. The reasoning accounts for the significant difference between specificity (0.843) and sensitivity (0.706) obtained through the CNN-LSTM model on the testing data set. Similarly, in age bias group 2, the normal participants are over-represented by a relatively older population (aged above 40 years). This leads the model to treat eldelry participants as normal participants which is again shown by the difference in the obtained specificity (0.825) and sensitivity (0.687). This clearly indicates that CNN-LSTM is prone to learn features directly associated with the underlying age bias which led to the overestimation of the COVID-19 detection performance. Unlike CNN-LSTM, RBF-Net is immune to the impact of biases in the training data sets. The results of the RBF-Net model demonstrate a major improvement over the CNN-LSTM model in all of the performance metrics when evaluated on unseen testing data set. An accuracy of 0.845 and 0.818 is achieved across both training groups, respectively, and at the same time, the difference between sensitivity and specificity is also diminished. Thus, it demonstrates that the RBF-Net is also suitable to alleviate the age bias in real-world COVID-19 detection applications.

### C. Performance on the Smoking Status Biased Training Scenario

The impact of underlying bias in the smoking status distribution on the CNN-LSTM model causing the inaccurately inflating COVID-19 detection performance is being analyzed. Using 10-fold CV technique, CNN-LSTM model obtained an average accuracy of 0.862 as shown in Table IV; whereas, when evaluated on unseen testing data, the accuracy drops to 0.723 which is almost 14% less than the CV technique. This clearly shows that CNN-LSTM model learned features directly associated with the smoking status bias and its COVID-19 detection performance is again overestimated. On the contrary, RBF-Net shows its ability to eliminate the underlying impact of smoking status bias in the training scenario when it is evaluated on the unseen smoking status testing data set. The results of the RBF-Net model show a major improvement over the CNN-LSTM model in all of the performance metrics. The obtained accuracy, F1-score, and ROC-AUC is almost 8% higher compared to the CNN-LSTM model.

### D. Ablation Study

To further elucidate the novelty of our proposed approach, which involves integrating the bias predictor module (B) into the CNN-LSTM framework (module A) to address confounding variables, we conducted an additional experiment. In preceding sections, we have methodically demonstrated how our RBF-Net (A + B) consistently outperforms the standalone CNN-LSTM model when trained on diverse sets of biased data. In this ablation study, we aim to assess the specific influence of the bias predictor module (B) within the RBF-Net in comparison to the CNN and CNN-LSTM framework (A) under both unbiased and biased training conditions. The architecture used for constructing the CNN model is illustrated in [27]. To achieve this, we created a new training dataset devoid of any inherent bias related to the confounding variables of age, gender, and smoking status. This dataset consists of 900 samples for each target class, i.e., COVID-19 and normal. The gender distribution, age distribution, and smoking status distribution have roughly been kept the same across both of these classes; thus, eliminating any form of bias. Furthermore, we curated an additional balanced unseen testing dataset having 100 samples in both of the classes for evaluating the RBF-Net, CNN, and CNN-LSTM on the unbiased training data. In Table V below, we present the testing performance metrics for the CNN model, CNN-LSTM model (Module A) and the RBF-Net (A + B) when trained on the unbiased, gender biased, age biased (group 1), and smoking status biased training datasets and evaluated on their respective unseen testing datasets. This presentation is designed to highlight the discernible impact of the bias predictor (Module B) on the performance of RBF-Net.

TABLE V: Results of Ablation Studies on RBF-Net

| Model | Unbiased | | | Gender Biased | | | Age Biased | | | Smoking Status Biased | | |
|---|---|---|---|---|---|---|---|---|---|---|---|---|
| | Acc | F1-Score | ROC-AUC | Acc | F1-Score | ROC-AUC | Acc | F1-Score | ROC-AUC | Acc | F1-Score | F1-Score |
| CNN (baseline) | 0.821 | 0.822 | 0.822 | 0.751 | 0.751 | 0.751 | 0.746 | 0.741 | 0.746 | 0.683 | 0.679 | 0.684 |
| CNN-LSTM (A) | 0.874 | 0.877 | 0.874 | 0.787 | 0.785 | 0.787 | 0.774 | 0.775 | 0.776 | 0.723 | 0.727 | 0.723 |
| RBF-Net (A+B) | 0.879 | 0.883 | 0.878 | 0.841 | 0.845 | 0.846 | 0.845 | 0.845 | 0.846 | 0.805 | 0.811 | 0.805 |

Feature Encoder module is a shared component in both CNN-LSTM and RBF-Net, which plays the fundamental role in learning the distinctive features influenced by the presence of COVID-19. For this reason, the performance achieved by CNN-LSTM and RBF-Net is similar in the unbiased training setting. However, the vitality of the bias predictor (module B) in the RBF-Net is established when both models are trained and evaluated under biased training conditions. In these scenarios, the performance of the CNN-LSTM model experiences a significant decline, whereas the RBF-Net remains resilient, consistently maintaining its accuracy, as highlighted in the preceding sections. Consequently, the incorporation of the bias predictor module (module B) within the c-GAN framework significantly bolsters the RBF-Net's capability to learn the nuanced impact of COVID-19 features. This enhancement not only contributes to its effectiveness but also renders it more practical for deployment as a digital testing tool.

## VI. Discussion and limitations

As discussed in the preceding sections, our investigation delved into the influence of three pivotal confounding variables: gender, age, and smoking status. These variables were meticulously considered to assess their impact on the CNN-LSTM model's ability to detect COVID-19. Notably, the substantial performance degradation observed in the CNN-LSTM model—from the performance achieved during cross-validation to its subsequent evaluation on the balanced testing data—clearly demonstrates the adverse effects of data selection bias on the features learned by the model. Thus, the underlying demographic bias, that often exists in the distributions of real-world clinical data [34], [35], [39], has to be addressed in future machine learning schemes, thereby making them practical and effective tools for validation by healthcare practitioners and clinicians in the realm of digital healthcare solutions.

The RBF-Net model effectively mitigates the influence of inherent biases present in training data distributions by extracting meaningful features from them. This is vividly illustrated by the significant performance improvements exhibited by the RBF-Net framework when compared to the CNN-LSTM model across all confounding variables. The differences in performance metrics, including accuracy, specificity, sensitivity, F1-score, and ROC-AUC, across various biased training groups are depicted in Fig. 7. A noteworthy insight from the results is that the RBF-Net model achieved the most substantial improvement when dealing with the smoking status biased training group. It achieved an approximately 8% increase in accuracy, F1-score, and ROC-AUC in this specific scenario. The architecture of the feature encoder module within the RBF-Net sheds light on the reason behind this remarkable improvement. The encoder comprises convolutional blocks and an additional LSTM block, which collectively identify features that hold spatial and temporal significance. This feature extraction capability enables the RBF-Net model to effectively discern the impact of COVID-19 on cough spectrogram images, as COVID-19 often manifests respiratory symptoms that, in turn, influence the spatio-temporal features of these cough spectrograms, as observed in previous studies [10]. Simultaneously, smoking also induces changes in human cough spectrograms in the spatio-temporal domain. This implies that the effect of smoking on the spectrograms bears some resemblance to the impact of COVID-19, especially when contrasted with the effects of gender and age on the spectrograms. Consequently, the RBF-Net framework yields the maximum performance increase when mitigating the impact of the bias related to smoking status.

Another standout aspect of our study lies in the high-quality data used for training and testing the DL model, enhancing the credibility of our results and conclusions. Unlike many existing cough-acoustic datasets [27], [29], [35], [48], [49], which often rely on crowd-sourced data, the data used n this paper was meticulously collected within a reputable medical facility under the supervision of trained healthcare professionals. Furthermore, the cough audio samples were obtained on the same day as the COVID-19 labels were assigned through standardized RT-PCR tests. This meticulous approach minimizes the potential errors stemming from misreporting or participants' lack of awareness regarding their COVID-19 status at the time of cough recording.

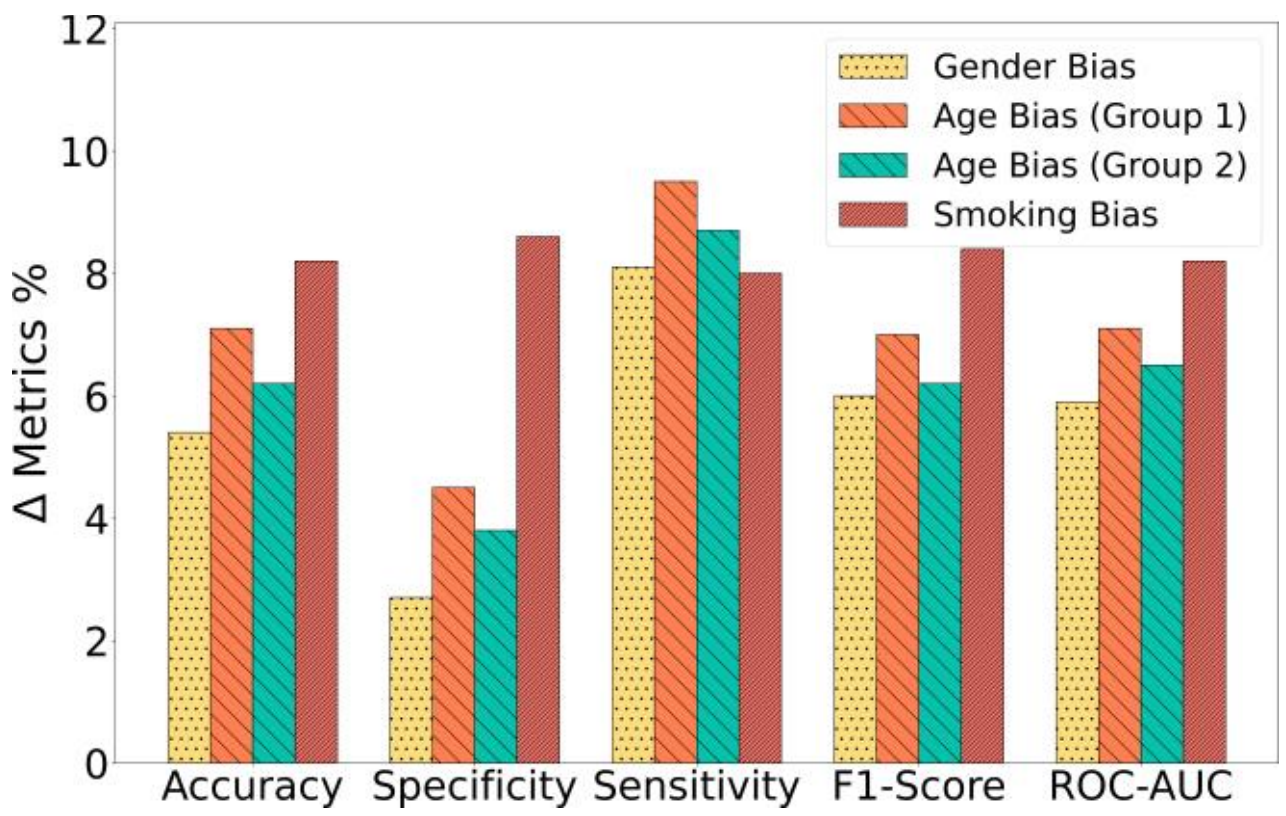

Fig. 7: Improvement in the performance metrics (accuracy, specificity, sensitivity, F1-score, and ROC-AUC) achieved by the RBF-Net compared to the CNN-LSTM.

One limitation of our current scope of work is that the efficacy of the RBF-Net framework has thus far been validated solely on the COVID-19 cough dataset and has not yet been extended to other RDs such as tuberculosis, asthma, and COPD. The data collection process for these additional conditions is currently underway within a medical facility, a process that demands a considerable amount of time and resources, while adhering to strict Institutional Review Board (IRB) protocols. Additionally, ensuring patient privacy, obtaining ground-truth diagnosis information via gold standard tests at the precise moment of cough sound collection, and maintaining consistency in the cough sound sample collection by minimizing device variability, background noise, and other environmental factors are integral to this endeavor. Our immediate future agenda revolves around evaluating RBF-Net's capability to identify respiratory diseases akin to COVID-19 based on the acoustic signatures present in cough sounds. This assessment will be conducted once the data collection efforts are successfully concluded.

Our upcoming research, will have a bi-fold objective. Firstly, we intend to broaden the spectrum of confounding variables by incorporating the common influenza virus, a respiratory tract ailment. This addition will allow us to explore how the presence of the influenza virus might impact the performance of our RBF-Net in COVID-19 and other RDs detection. By doing so, we can unravel potential interactions between these respiratory conditions, thus improving the accuracy of our models. Secondly, we plan to delve into the influence of other common confounding factors, such as race and geographical location. These factors can play a pivotal role in shaping the prevalence of various diseases and introduce biases in data distributions. Examining and accounting for these variables will not only bolster the resilience of the RBF-Net framework but will also make it adaptable for global-scale screening efforts. This comprehensive approach will enhance our understanding of the interplay between various confounding factors and the performance of our model, ultimately facilitating more accurate and globally relevant disease detection and screening.

## VII. Conclusion

This paper addresses the challenge of mitigating underlying bias in training data distributions, which often leads to inflated respiratory disease (RD) diagnosis results. We introduce modifications to the CNN-LSTM architecture to formulate the RBF-Net framework, incorporating an additional bias predictor module. This module assesses the statistical association between the feature vector and biases, aiming to reduce correlations between confounding variables and RD classification outcomes. To illustrate the impact of underlying biases on DL model performance and evaluate the RBF-Net's effectiveness, we utilize multiple training datasets with various biases (gender, age, smoking status), along with balanced unseen testing datasets. Both the state-of-the-art CNN-LSTM and the proposed RBF-Net models undergo training and evaluation on these datasets. Notably, the RBF-Net framework demonstrates promising and realistic results, even in challenging biased training scenarios, and without relying on cross-validation techniques. These findings suggest the potential of the proposed framework as an effective non-invasive tool for RD testing and screening.

**Author Contributions:** Conceptualization, T. S., A. I., and A. I., A. R.; methodology, T. S. and A. I.; software, T. S.; validation, T. S. and A. I., A. R; formal analysis, T. S.; investigation, T. S. and A. I., A. I., A. R.; resources, T. S., A. I., A. R., and A. I.; data curation, T. S. and A. I.; writing draft preparation, T. S. and A. I; writing—review and editing, T. S., A. I, I. S., H. N. Q., A. R., and A. I.; visualization, T. S. and A. I. All authors have read and agreed to the published version of the manuscript.

**Declaration of competing Interest:** The authors declare that they have no known competing conflict of interests or personal relationships that could have appeared to influence the work reported in this paper.

**Data Availability Statement:** The statistical data presented in this study are available in Fig. 1, 3, 4, and 5. The data sets used and/or analyzed during the current study can be available upon request. These data are not publicly available due to privacy and ethical reasons.

**Acknowledgments:** This work was not supported by any grant.

**Conflicts of Interest:** The authors declare no conflict of interest.